\documentclass[journal]{IEEEtran}

\usepackage{color}
\usepackage[pdftex,bookmarks=false]{hyperref}
\usepackage{graphicx} \usepackage{subfigure} \usepackage{url} 
\usepackage{amsmath}
\usepackage{algorithm}
\usepackage{algpseudocode}
\usepackage{pifont}
\usepackage{amsfonts}
\usepackage{balance}
\usepackage{cite}
\usepackage{multirow}
\usepackage{array,booktabs}
\usepackage{tabularx}
\usepackage{rotating}

\usepackage{paralist}
\usepackage{authblk}
\usepackage{steinmetz}
\usepackage{epstopdf}
\usepackage{tablefootnote}

\usepackage{amsthm}
\theoremstyle{plain}

\DeclareMathAlphabet{\mathpzc}{OT1}{pzc}{m}{it}

\usepackage[inline]{LatexPackage/trackchanges}
\begin{document}

\title{Towards a Heterogeneous Smart Electromagnetic Environment for Millimeter-Wave Communications: An Industrial Viewpoint}

\author{Roberto~Flamini,
				Danilo~De~Donno,
				Jonathan~Gambini,
				Francesco~Giuppi,
				Christian~Mazzucco,
				Angelo~Milani,
				and Laura~Resteghini
\thanks{Manuscript received October xx, 2021.}
\thanks{The authors are with the Milan Research Center of Huawei Technologies Italia S.r.l, Milan, Italy, (e-mail: name.surname@huawei.com).}}

\markboth{IEEE Transactions on Antennas and Propagation}%
{Shell \MakeLowercase{\textit{et al.}}}

\maketitle

\newcommand{\ed}{\color{blue}}
\newcommand{\edd}{\color{red}}
\newcommand{\PRF}{\textbf{P}_{\text{RF}}}
\newcommand{\PBB}{\textbf{P}_{\text{BB}}}
\newcommand{\CRF}{\textbf{C}_{\text{RF}}}
\newcommand{\CBB}{\textbf{C\emph{}}_{\text{BB}}}
\newcommand{\VN}[2]{{\textbf{#1}_{\text{#2}}}}
\newcommand{\SN}[2]{{{#1}_{\text{#2}}}}
\newcommand{\specialcell}[2][c]{%
  \begin{tabular}[#1]{@{}c@{}}#2\end{tabular}}
	\newcolumntype{M}[1]{>{\centering\arraybackslash}m{#1}}
\newcolumntype{C}{>{\raggedright\arraybackslash}X}
\newcolumntype{Y}{>{\centering\arraybackslash}X}

\makeatletter
\newcommand\footnoteref[1]{\protected@xdef\@thefnmark{\ref{#1}}\@footnotemark}
\makeatother

\bstctlcite{mybib:BSTcontrol}

\begin{abstract}

Fifth generation (5G) and beyond communication systems open the door to millimeter Wave (mmWave) frequency bands to leverage the extremely large operating bandwidths and deliver unprecedented network capacity. These frequency bands are affected by high propagation losses that severely limit the achievable coverage. The simplest way to address this problem would be to increase the number of installed mmWave base stations (BSs), at the same time augmenting the overall network cost, power consumption and ElectroMagnetic Field (EMF) levels. As alternative direction, here we propose to complement the macro-layer of mmWave BSs with a heterogeneous deployment of Smart Electromagnetic (Smart EM) Entities -- namely IAB nodes, Smart Repeaters, Reconfigurable Intelligent Surfaces (RISs) and passive surfaces -- that is judiciously planned to minimize the total installation costs while at the same time optimizing the network spectral efficiency. Initial network planning results underline the effectiveness of the proposed approach. The available technologies and the key research directions for achieving this view are thoroughly discussed by accounting for issues ranging from system-level design to the development of new materials.

\end{abstract}


\IEEEpeerreviewmaketitle

\section{Introduction}

\IEEEPARstart{T}{he} 5G New Radio (NR) is the first mobile technology generation to make use of the underutilized mmWave spectrum between 24.25 GHz and 52.6 GHz, which has been defined by the 3GPP standard as Frequency Range 2 (FR2) \cite{3GPP_5G_NR_1,3GPP_5G_NR_2}. At the same time, the large amount of bandwidth availability in the unlicensed mmWave band at 60 GHz has led to the formulation of next generation WiFi standards like IEEE 802.11ad and 802.11ay, also known as Wireless Gigabit (WiGig) \cite{802_11_ay}. The mmWave spectrum is extremely attractive for two main reasons: (\textit{i}) it allows to accommodate large antenna arrays in very compact systems, thus obtaining high antenna gain and beamforming capability; and (\textit{ii}) it provides large channel bandwidths. These points make mmWave suitable to accommodate a rapidly growing number of intelligent devices and services and to address highly demanding applications, where very low latencies and very high data rates are required \cite{paper_lombardi, 1-5G}.

As a critical aspect, communications in the mmWave frequency range are essentially coverage-limited if compared with sub-6GHz and this limitation represents the most important issue that mobile operators, ICT manufacturers and standardization bodies are currently facing in order to be able to successfully plan and deploy 5G networks operating at FR2. Coverage limitation at mmWave is much more severe than at sub-6GHz because of two main reasons: (\textit{i}) the higher free space path loss (FSPL) and (\textit{ii}) the higher sensitivity to obstacles in the propagation environment. In terms of FSPL, it is sufficient to consider that a radio link at 28GHz suffers from a loss of around 29dB higher than that at 1GHz and this dramatically affects the link budget. Regarding the role of the environment, mmWave signals are subject to higher attenuation in presence of blockage from obstacles of different kinds, including the human body, foliage and building walls. Moreover, due to the small wavelength, the roughness and imperfections of the surfaces tend to create a scattering effect rather than a specular theoretical reflection, leading to a weaker signal strength \cite{book_rappaport}.

To mitigate these limitations, mmWave applications typically resort to high gain and narrow beamforming antenna arrays combined with high transmit power, in other words high EIRP systems. While this solution is sufficient for adequately recovering the propagation losses in line-of-sight (LoS) or near-LoS scenarios, it fails in delivering acceptable performance in urban areas characterized by deep non-LoS (NLoS) conditions. For this cases, a natural solution would be to resort to network densification. A denser deployment of 5G mmWave base stations (gNBs according to the 5G NR terminology) would be preferred from the performance point of view since it would guarantee the desired minimum signal strength in the served area, but it may not be always a feasible or economically viable solution, e.g., due to the lack of wired/wireless backhaul, the higher costs for the acquisition of new sites, rental fees, maintenance and power supply. As a matter of fact, alternative solutions to the approach ''more information and data through more power and higher costs'' are mandatory. Aligned to this view, standardization bodies, mobile operators and ICT industries are now evaluating a number of trade-off solutions in terms of network performance and complexity, total cost of ownership (TCO) and electromagnetic field (EMF) levels. Such solutions include, among the others, the deployment of new types of network nodes, much cheaper and easy to be deployed than traditional Macro and Micro BSs, that allow to achieve an homogeneous coverage in the radio access network without necessarily resorting to the installation of additional macro-sites.

\begin{figure}[t]
\centering
\includegraphics[width=0.48\textwidth]{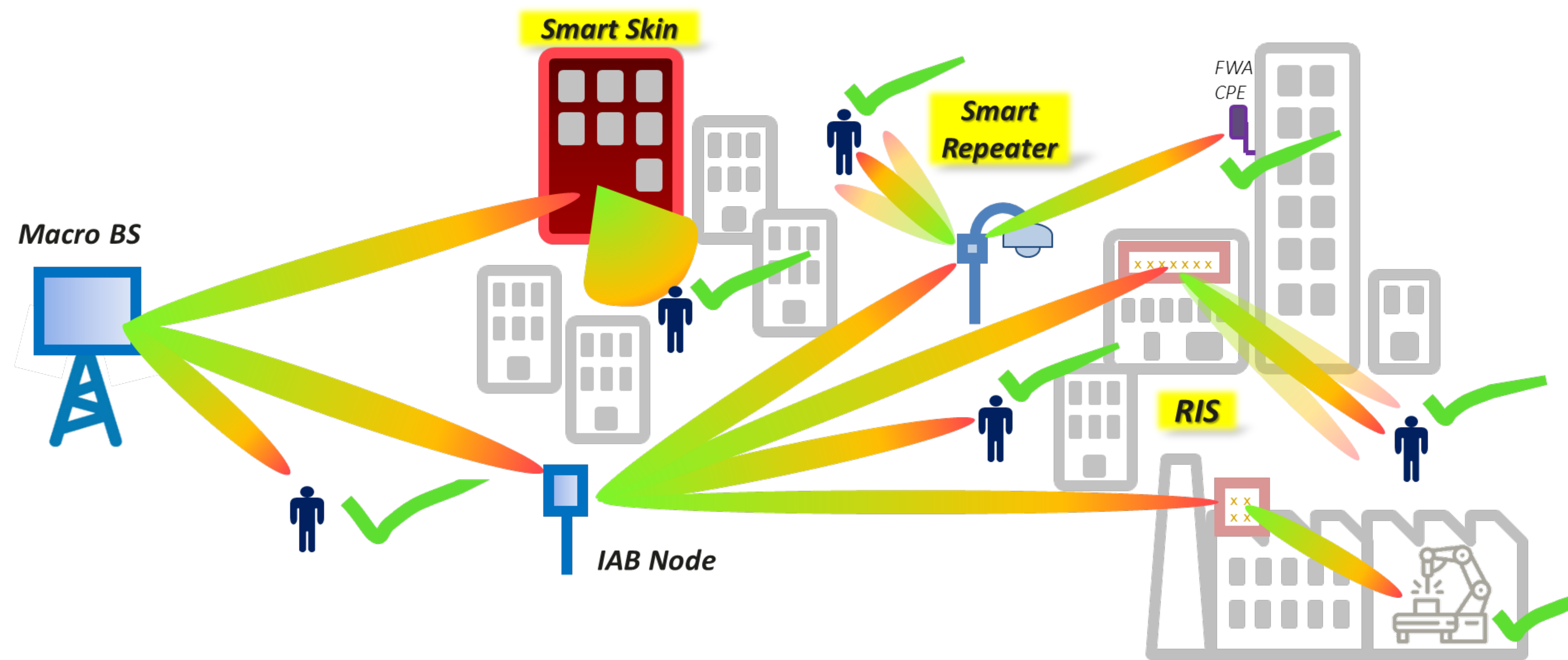}
\caption{Smart EM Environment: an environment leveraging the electromagnetic properties of new network nodes, whose deployment and operation is optimized to maximize network performance in terms of coverage and capacity while reducing costs and E2E power consumption.}
\label{fig:smartem_env}
\end{figure}

\noindent \textit{\textbf{Contribution.}} In this paper, we focus on the concept of a Heterogeneous Smart Electromagnetic (EM) Environment as a new paradigm where novel types of network nodes -- specifically, passive static surfaces, Reconfigurable Surfaces, Smart Repeaters, and Integrated Access and Backahul (IAB) nodes -- coexist in the same network and are optimized in terms of deployment and functionality to maximize network performance (e.g., coverage and capacity) while reducing costs and end-to-end power consumption. In Fig.~\ref{fig:smartem_env}, we sketch our vision of Smart EM Environment. In this framework, we propose a system-level study to show the advantages of the outlined solution with respect to more conservative network planning approaches. Moreover, we clearly identify the main technological directions as well as the most promising design approaches that can be successfully taken into account for an effective development, manufacturing and deployment of the new Smart EM Entities. To this end, we provide an inspiring review of the most interesting tunable materials suitable for reconfigurable antennas and passive/active surfaces. 

The paper is organized as follows. In \S \ref{actors}, a detailed description of the network nodes forming a Smart EM Environment is provided.  In \S \ref{applications}, an overview of the potential application scenarios for the Smart
EM Environment at mmWave frequencies is given, along with some updates on industrial progress and standardization activities related to this topic. The concept of heterogeneous deployment is analyzed in \S \ref{het_deployment} where preliminary system-level evaluations are presented to show the advantages of the proposed solution. Finally, the main system-level and hardware-level challenges, as well as the most promising design approaches to manufacture and deploy Smart EM nodes are outlined in \S \ref{system_challenges} and \S \ref{hardware_challenges}, respectively.

\section{Smart EM Environment: the actors} \label{actors}

\begin{figure*}[t]
 \centering
 \includegraphics[width=1.0\textwidth]{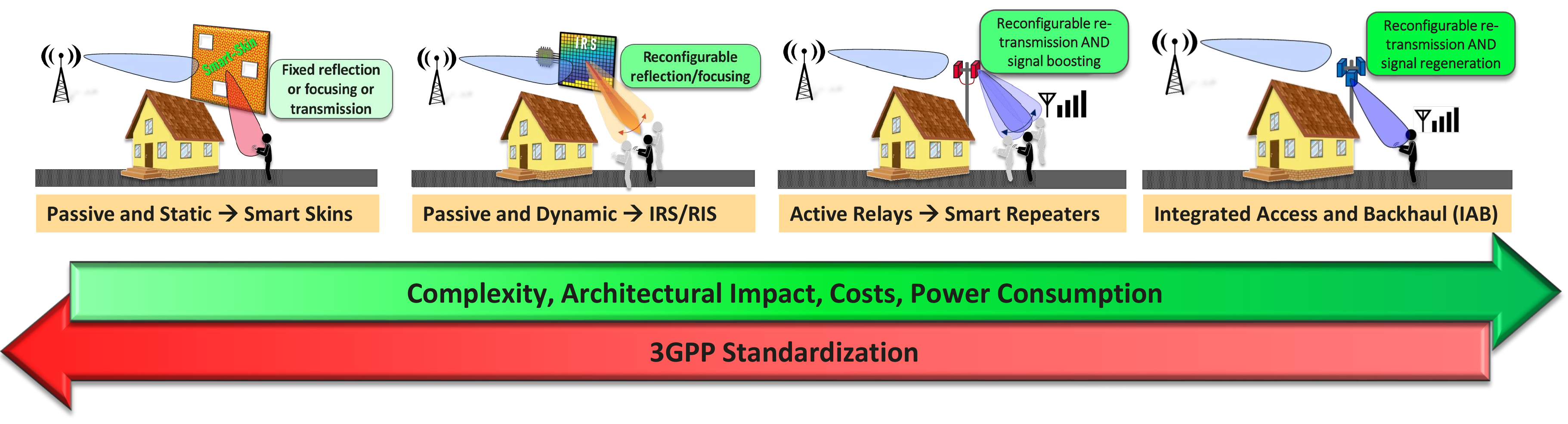} 
\caption{The actors forming a Smart EM Environment presented, from left to right, in increasing order of complexity, architectural impact, cost, power consumption and, from right to left, in terms of timeline for 3GPP standardization.} 
	\label{fig:actors}
\end{figure*}

The use of higher frequencies, especially -- but not only -- in the mmWave spectrum, is extremely attractive in terms of available bandwidth, but must cope with higher propagation losses with respect to the sub-6GHz bands. For this reason, new types of network nodes have been considered in order to increase deployment flexibility, that at the end translates into new paradigms for mobile operators to design and operate their networks. In this section, we take Fig.~\ref{fig:actors} as a reference to introduce the four types of network nodes envisioned for the Smart EM Environment in order of decreasing complexity, cost, power consumption, and architectural impact. The main characteristics are also summarized in Table~\ref{tab:parameters}.

\subsection{IAB node}
As per 3GPP specifications, an IAB Radio Access Network (RAN) generally consists of a network of micro base stations (IAB nodes) interconnected through in-band wireless backhaul to a Macro BS (or Donor gNB) that, instead, has direct, wired access to the core network. Consequently, IAB brings about affordable dense networks without the need to densify backhaul wiring \cite{3gpp_iab}. Moreover, IAB is particularly effective when applied to mmWave RAN: thanks to the vast wireless capacity and highly directional transmissions, it is possible to obtain dense access networks by installing only a few pricey Donor gNBs.

The peculiar aspect of IAB is the self-backhauling approach, where both radio access and wireless backhaul links share the same radio resources and interfaces. Therefore, a proper management of the radio resource allocation is fundamental to operate this network and it is carried out by the Donor gNB. The key benefits of IAB are essentially three: 

\begin{enumerate}
	\item it provides fast mechanisms to handle user mobility and handover without directly involving the core network;
	\item it guarantees coverage extension for incremental network deployment;
	\item it enhances overall cell capacity and cell-edge user experience. 
\end{enumerate}

The Donor gNB consists of a central unit (CU) and one or more distributed units (DUs). A CU is a logical node hosting radio resource control (RRC), service data adaptation protocol (SDAP) and packet data convergence protocol (PDCP) and controls the operation of one or more DUs. A DU is a logical node hosting radio link control (RLC), medium access control (MAC) and physical (PHY) layers. IAB nodes exploit the CU/DU functional split by hosting a DU and a \textit{mobile termination} (MT) function. The IAB-DU interfaces to the Donor gNB CU via the F1 interface (with relevant extensions). The MT replicates the UE functionality to establish connections to a gNB and the core network. In this manner, IAB-MTs can connect to DUs on other IAB nodes or on the Donor gNB thus realizing multi-hop backhaul. The CU on the Donor gNB becomes the central control function for all IAB-DUs and IAB-MTs defining an interconnected IAB topology. 

Based on the above discussion, it is clear that IAB nodes are Layer-2 (L2) regenerative relays, as every packet received (either from the Donor gNB or from the UE or from another IAB Node) has to be properly decoded and re-encoded before re-transmission. IAB has been standardized in 3GPP Rel-16 \cite{IAB_standard} assuming half-duplex operation between access and backhaul transmission and reception. Nevertheless, the plan is to evolve towards full-duplex operation in 3GPP Rel-17 with significant expected advantages in terms of network sum capacity.

From the hardware point of view, the IAB-MT could either use a separate antenna
or share the access antenna used by the IAB-DU (virtual IAB-MT). The latter provides the ultimate level of integration and allows to use the high-performance base station antennas for backhaul over longer distances. On the access part, IAB nodes are a sort of access points with compact form factor, approximately with 8 Kg weight, 50-60dBm Equivalent Isotropically Radiated Power (EIRP), and 300-350W power consumption.

\subsection{Smart Repeater}
The Smart Repeater will be introduced in 3GPP Rel-18 under the name of Network-Controlled Repeater \cite{3GPP_NCR} as an evolution of the more classical RF Repeater, which has been used in 2G, 3G and 4G deployments to supplement the coverage provided by regular macro cells. Classical RF Repeaters are low-cost, easy to deploy, and do not increase latency. On the other hand, as they amplify both signal and noise, they can contribute to an increase of interference in the system. RF repeaters are also non-regenerative type of relay nodes, as they do not decode and re-encode the signal as done by IAB nodes, but simply amplify-and-forward the received signal. Moreover, they are full-duplex nodes, not able to differentiate between uplink and downlink, and usually not designed to provide beamforming gain.

The concept of Smart Repeater has been proposed within the 3GPP as a new type of network node that is able to make use of some side control information to enable a more intelligent amplify-and-forward operation in a system with time-division duplex (TDD) access and beamforming operation \cite{qualcomm_rel_18}. The Smart Repeater is still non-regenerative and requires a low-capacity control channel from the Donor gNB. From the hardware perspective a Smart Repeater can be thought as consisting of two beamforming antennas (e.g., phased antenna arrays), one oriented towards the serving Donor gNB while the other oriented towards the service area to be covered. The two antennas allows to achieve the highest flexibility in terms of installation and therefore in terms of size/shape/location of the target area. A Smart Repeater with higher level of integration (e.g., with the two antennas arranged back-to-back in a single box) can be envisioned to further reduce costs, dimensions, and to have a limited impact on urban architecture. Generally speaking, we imagine a Smart Repeater with small form factor, light weight (roughly 4Kg), with less than 60dBm EIRP, and low power consumption ($\sim$20W).

\subsection{Reconfigurable Intelligent Surface (RIS)}
Reconfigurable Intelligent Surface (RIS) is a new type of system node with reconfigurable surface technology, where its response can be adapted to the status of the propagation environment through control signaling, hence turning the wireless environment from a passive to an intelligent actor as the channel becomes somehow programmable \cite{bjornson2019intelligent, Wang2019, di2020reconfigurable, Renzo2020}. An Industry Specification Group (ISG) has been recently created within the ETSI (European Telecommunications Standards Institute) organization with the aim to study and specify use cases, deployment scenarios, evaluation methods, system architecture and requirements of RIS as well as its impact on the current specifications published by Standards Development Organizations (SDOs) like the ETSI or 3GPP. A RIS can be defined as a semi-passive or nearly-passive device since it does not include any amplification stage and requires relatively low power to be configured and controlled. For these reasons, RISs (\textit{i}) do not amplify nor introduce noise when reflecting/transmitting/reshaping the signals, (\textit{ii}) do not increase the latency providing an inherently full-duplex transmission, and (\textit{iii}) do not increase the EMF levels. From the hardware point of view, an RIS can be seen as a planar surface composed by an arrangement of a massive amount of controllable unit cells \cite{ris_garcia} -- whose size is in the order of $\lambda/2$ (being $\lambda$ the wavelength) in case of reflectarray-like design or smaller fractions of $\lambda$ for meta-material based approaches -- that are able to capture and re-radiate energy in a controllable manner \cite{di2020reconfigurable,Renzo2020, gacanin2020wireless}.

RISs are designed to be installed on walls, building facades, ceilings or even hung on lamp posts, poles or towers, generally in Line of Sight (LoS) with the serving Donor gNB (or IAB node or Smart Repeater or another RIS). The size (area) of a RIS must be chosen as the best compromise between coverage improvement, cost, power consumption and control complexity, which in turn depends on how fast the RIS configuration needs to be updated. Tipically, surface sizes of 25x25cm$^2$ or 50x50cm$^2$ are suitable for mmWave applications, while overall $<$2W power consumption can be considered as a target to make RIS a valid, low-cost alternative to Smart Repeaters.

\begin{table*}[]
\centering 
\caption{Overview of the main characteristics and parameters of Smart EM devices}
\begin{tabular}{m{2.5cm}|m{3.5cm}|m{5cm}|m{5cm}}
\hline \hline
Network node & \centering Brief description & \centering Main parameters & \centering Configuration  \tabularnewline \hline 
IAB Node &  L2 regenerative relay which can decode and forward RF signals & Relatively small form factor, around 8Kg weight, 60dBm EIRP, $\leq$350W power consumption & 5G control plane, real-time beamforming at TTI level \\ \cline{1-4}  
Smart Repeater & Non-regenerative relay which can amplify and forward RF signals & Small form factor, lightweight (around 4Kg), $<$60dBm EIRP, $\sim$20W power consumption & \multirow{2}{\hsize}{Two possible modes of operation: (i) semi-dynamic with out-of-band control plane and small number of beams updated few times per day; (ii) dynamic with in-band control plane and real-time beam update at TTI level.} \\ &&&\\ \cline{1-3} 
RIS & Intelligent radio surface whose response can be reconfigured using control signalling & Tipically 25x25cm$^2$ or 50x50cm$^2$ surface dimension for mmWave applications, $<$2W power consumption. & \\ \hline
Smart Skin & Surface that can be appropriately engineered/designed to go beyond Snell's law, providing, for instance, non-specular reflection and beam-shaping & $>$25x25cm$^2$ surface dimension, fully passive. & Fixed configuration \\ \cline{1-4}
Massive MIMO BS & Antenna-integrated radio, typically comprising many transceiver chains & Cell tower site, around 20Kg weight, up to 70dBm EIRP, $\leq$800W power consumption & Real-time beamforming at TTI level  \\ \hline \hline
\end{tabular}
\label{tab:parameters}
\end{table*}

\subsection{Smart Skin}
The fourth new type of network node is the Intelligent Surface (or Smart Surface, or Smart Skin). From now on along the paper we simply refer to it as Smart Skin. As shown in Fig.~\ref{fig:actors}, a Smart Skin is the simplest node in a Smart EM Environment, in the sense that it is a fully passive, non-reconfigurable surface able to reflect (or focus, or absorb, or even transmit) an impinging wave with a non-specular reflection angle. In addition to that, being non-reconfigurable, a Smart Skin does not require any control signaling and therefore it has no impact on the network operation, but only on network planning. From the hardware perspective, a Smart Skin can be thought as a surface, either planar or conformal, designed to be installed on walls, building facades, ceilings or even hung on lamp posts, poles or towers, in Line of Sight (LoS) with the serving Donor gNB (or IAB node or Smart Repeater or RIS). Being fully-passive and non-reconfigurable, the Smart Skin does need neither reconfigurable devices, nor power supply and therefore it can be fabricated with extremely low manufacturing costs. This characteristic allows the Smart Skin to be much larger in size than RISs or, equivalently, Smart Skins can be deployed more massively compared to RISs.

\section{Application scenarios, industrial progress and standardization} \label{applications}

\begin{figure}[t]
\centering
\includegraphics[width=0.47\textwidth]{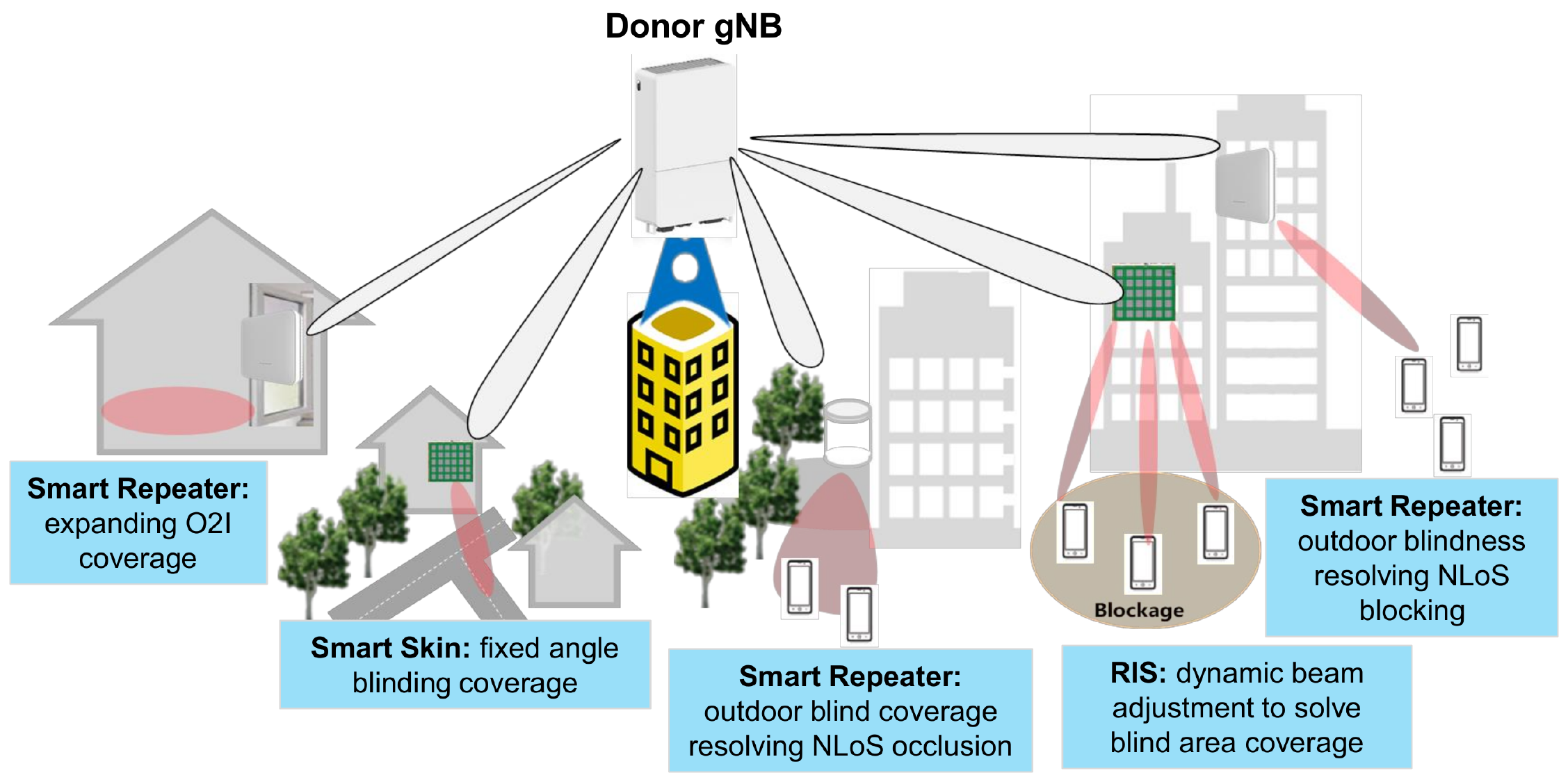}
\caption{Heterogeneous deployment for the Smart EM Environment.}
\label{fig:SEM_apps}
\end{figure}

This section is aimed at providing an industrial view on the potential applications and scenarios for the Smart EM Environment at mmWave frequencies along with a summary of industrial progress and standardization activities. A pictorial overview of Smart EM application scenarios is given in Fig.~\ref{fig:SEM_apps}. 

\subsection{Fixed Wireless Access (FWA)}
FWA brings high-speed broadband access to domestic and business sites. Users connect to the network by a Customer Premises Equipment (CPE) that in turn is served by a remote mmWave BS (placed at a distance of up to 2-3 km), typically over a LoS connection. In order to increase market penetration, FWA should guarantee the minimum Quality of Service (QoS) also in case of quasi-LOS (or even Non-LOS) conditions.  IAB nodes and Smart Repeaters with high end-to-end gain (typically set in the order of 90-100dB) appear as the best candidates to guarantee relevant spectral efficiencies in long distance regime.

\subsection{Enhanced Mobile Broadband (eMBB)}
eMBB is one of the three primary 5G New Radio (NR) use cases defined by the 3GPP. eMBB is supposed to provide broadband access to radio terminals in densely populated areas, both indoors and outdoors (like city centers, office buildings or public venues such as stadiums or conference centers), as well as mobile services to moving vehicles (including cars, buses, trains and planes). Typical urban and dense urban cell radii range from 200m to 500 m, and the target areas to be served are expected to include a high number of subscribers with frequent coverage problems and shadowing issues. In this framework, a heterogeneous deployment encompassing all the Smart EM actors (IAB Nodes, Smart Repeaters, RISs and Smart Skins) can be judiciously planned in order to strike an optimal balance among cost, capacity and coverage issues. 

%eMBB is actually one of three primary use cases 3GPP has defined for 5G NR with the objective to provide broadband access in densely populated areas, both indoors and outdoors (like city centers, office buildings or public venues like stadiums or conference centers) as well as mobile broadband services in moving vehicles (including cars, buses, trains and planes). For this scenario, the cell radius is in the order of hundreds of meters (200-300m) including a high number of subscribers with very likely coverage problems and shadowing issues. In order to overcome both coverage and capacity issues, all the Smart EM entities can play an important role: 
%
%\begin{itemize}
	%\item IAB would provide coverage and capacity;
	%\item Smart Repeaters would provide coverage and partly capacity;
	%\item Smart Skins and RISs would be suitable to solve coverage issues, but not capacity (assuming that Smart Skin/RIS is in LoS with the BS and UE is in LoS with Smart Skin/RIS, but in NLoS with the BS)
%\end{itemize}

\subsection{Verticals}
Differently from the previous generations of mobile systems, 5G (and beyond) technologies also aim at defining communication scenarios that can successfully lead to a global industrial transformation in many vertical markets such as transportation, media, healthcare, manufacturing, automotive, energy, public safety, smart cities and much more. One of the main challenges in this extremely diversified landscape lies in guaranteeing a homogenous coverage in industrial sites characterized by the presence of obstacles and numerous metal objects. In this context, all the Smart EM actors are potential candidates for enabling the use cases of Flexible Production Lines, Connected Workers and Automated Guided Vehicles for Industry 4.0. Moreover, the different network nodes can be deployed within the industrial sites with reduced operating costs due to absence of rental fees. This can be an enabling factor for the newest Smart EM solutions (Smart Repeater, RIS and Smart Skin) that inherently trade lower cost of apparatuses and installations for a more pervasive deployment with respect to, e.g., Macro BSs or IAB systems. Finally, passive solutions such as Smart Skins can be installed with a higher degree of flexibility than in public urban scenarios where architectural constraints and site acquisition campaigns set stricter constraints. 

\subsection{Industrial progress and standardization}
The most recent 3GPP studies and contributions are now all focused on Release 18 (Rel-18) that will mark the mid-generation of 5G with commercialization expected around 2025 \cite{3gpp_5G_adv}: the new term coined to indicate this new set of specifications is ''5G-Advanced''. 
%The vision of Huawei on 5G-Advanced develops around four main points, mainly driven by the needs of eMBB and verticals \cite{huawei_rel_18}. First, 5G advanced should set itself as a goal to provide significant improvements for all 5G bands: sub-3GHz FDD bands, sub-7GHz TDD bands, and mmWave bands. A second aspect is related to uplink boosting, following the requirement of increasing uplink capacity and user experience. The required features include flexible spectrum access, enhanced UL-MIMO, flexible duplexing, and multi-path UE Relay. A third aspect is related to Extended Reality (XR) with the need to satisfy higher-definition, low latency and higher-interactivity XR applications. The fourth aspect concerns the need to support new capabilities for verticals, such as low-power and high-accuracy positioning (LPHAP) and passive IoT, as well as continuing the requirements of other verticals.
Smart EM Environment, among the others, has been identified as a potential solution for enabling 5G-Advanced. In \cite{3GPP_NCR}, an evolution of the simple RF repeater, namely Smart Repeater or Network-Controlled Repeater, is presented posing much emphasis on its integration with IAB: the two technologies can be deployed independently, but also combined together to complement each other. IAB nodes add cells to the RAN footprint, while Smart Repeaters enhance the cell footprint of an IAB-DU or gNB-DU \cite{qualcomm_rel_18}. An important aspect is also that while IAB nodes are managed by the Donor gNB CU, the operation of Smart Repeaters can be managed by the IAB-DU or Donor gNB DU: this separation of control provides easier implementation and allows independent evolution of IAB and Smart Repeater. Moreover, keeping the Smart Repeater under local IAB-DU/gNB-DU control provides better scalability. Once the Smart Repeater will be specified in the standard, the related control mechanism may be the stepping stone towards RIS. 

Multi-beam aware, adaptive repeaters for Rel-18 are proposed in \cite{rakuten_rel_18} where it is shown how repeater performance can be improved by introducing control information such as slot transmission status, TDD slot configuration, TX/RX beam information and adaptive repeater bandwidth. 

In \cite{kddi_rel_18}, Smart Repeaters and RISs with adaptive beamforming control capabilities towards the UE in FR2 are proposed. Both solutions can serve not only as countermeasures against dead spots or outdoor-to-indoor attenuation, but also as boosters of MIMO layers.

Finally, in \cite{sony_rel_18, china_unicom_rel_18}, the introduction of RISs in 5G-Advanced is promoted by stressing the need to design a proper channel and device model, identify application scenarios, and analyze potential impacts on beam management, control interface with gNB, random access, and interference coordination.

\section{A promising solution: heterogeneous deployment} \label{het_deployment}

The issue of blockage due to opaque obstacles interrupting the line of sight is the most detrimental effect of mmWave propagation. A denser deployment of gNBs could provide the reliability level needed by 5G requirements, as failed radio links can be easily recovered by connecting to alternative gNBs. While such density increase might generate high deployment costs, the IAB paradigm outlined in the previous sections promises to limit the increased complexity and expenses, and it is thus considered as one among the most promising enablers of mmWave deployment. 
\begin{figure*}[t]
 \centering
 \subfigure[Coverage for gNB-only deployment]
   {\includegraphics[width=0.345\textwidth]{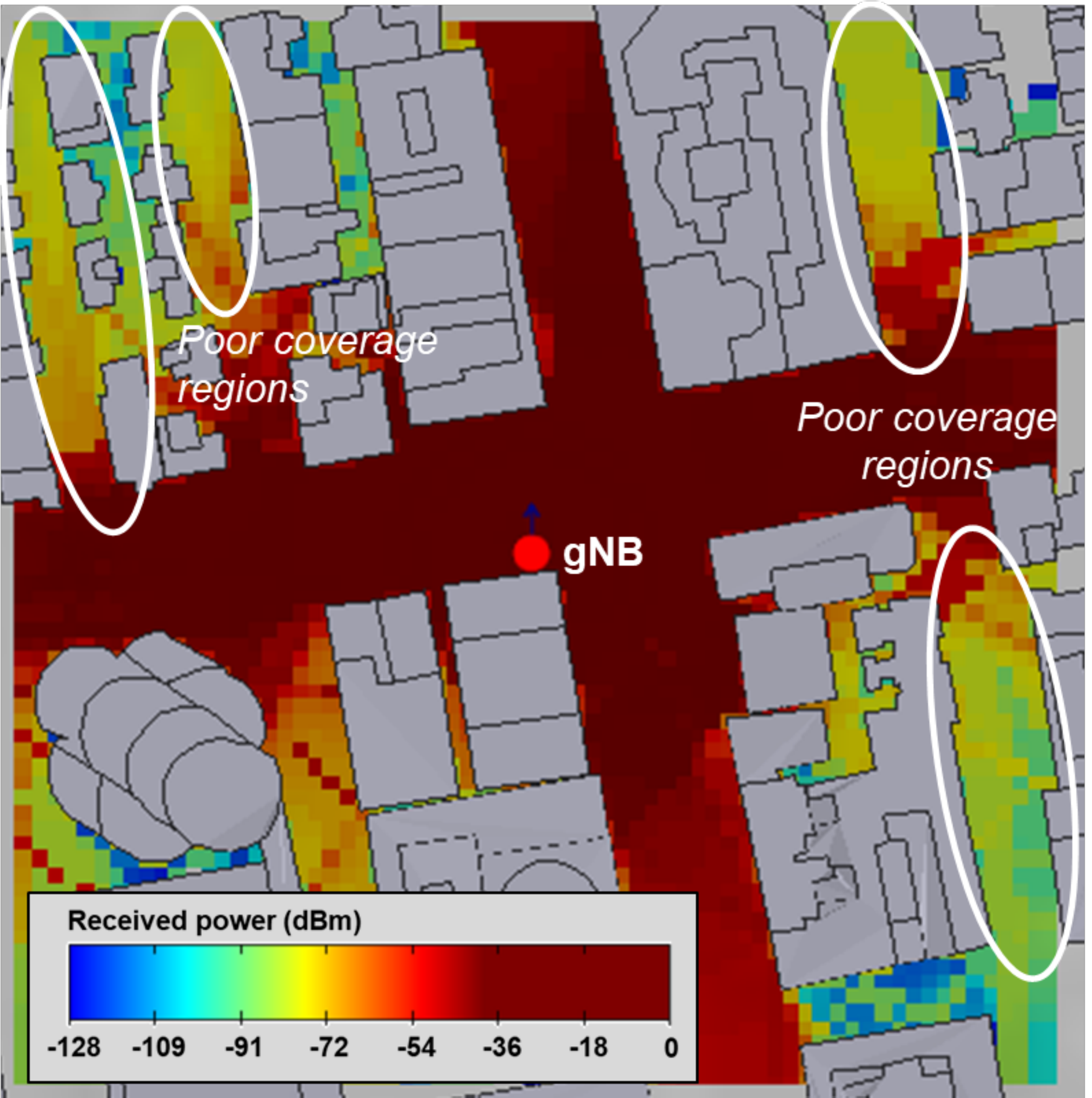} \label{fig:macro_only}}
 \hspace{20mm}
  \subfigure[Coverage for gNB+RISs deployment]
   {\includegraphics[width=0.35\textwidth]{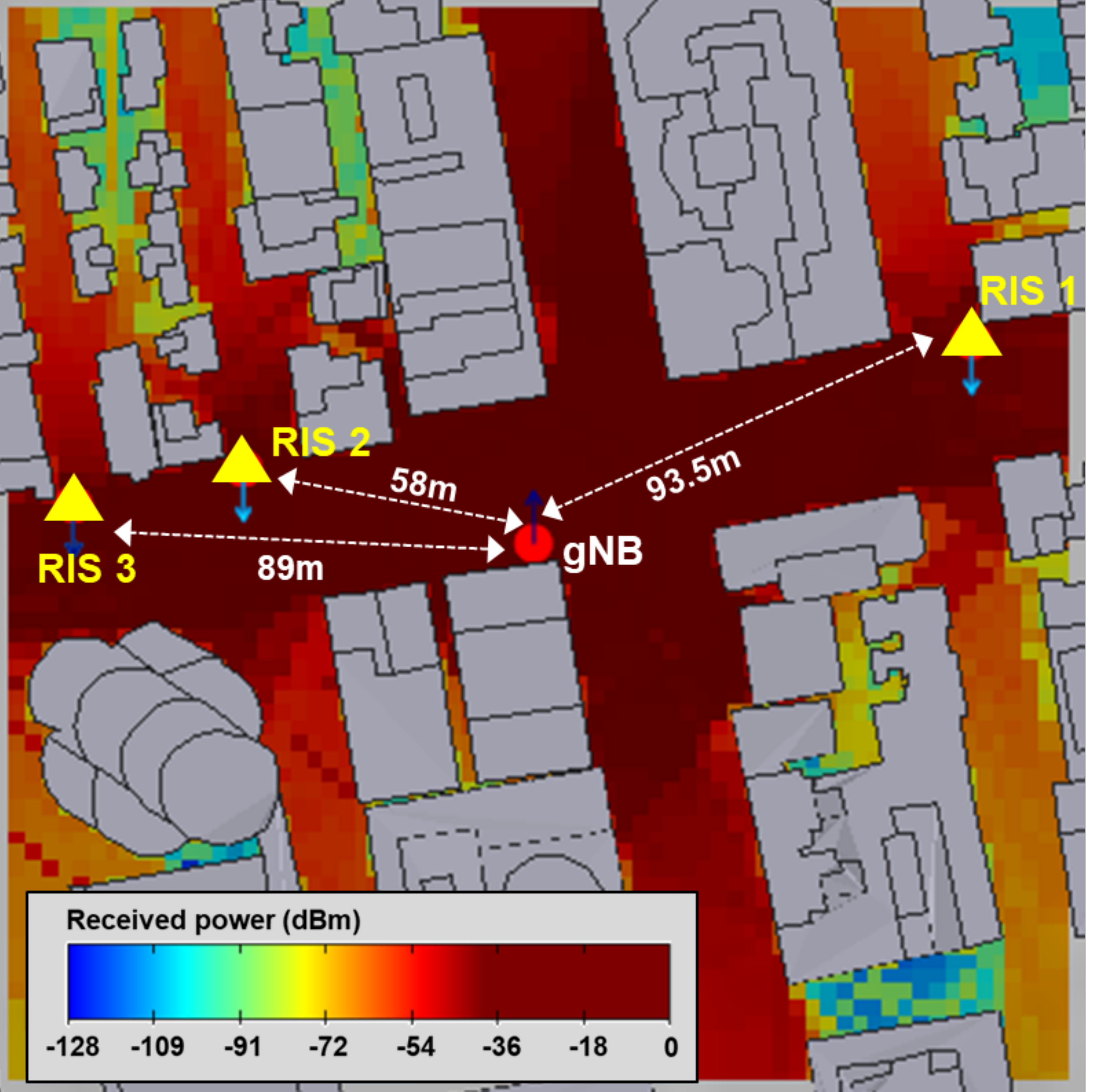} \label{fig:MACRO_RIS}}
 \caption{An example of heterogeneous deployment in a small area of Hong Kong city. Ray-tracing prediction shows significant coverage improvement when RISs are installed at strategic positions to fill 5G service holes.}
	\label{fig:HK_RIS_deployment}
 \end{figure*}

However, as already discussed, IAB nodes are full-fledged devices (i.e., they consists of DU and MT entities) compared to the other types of relay nodes for coverage augmentation presented in \S \ref{actors}, namely amplify-and-forward relays (Smart Repeaters), passive static reflectors (Smart Skins), and reconfigurable intelligent surfaces (RISs). In our view, these nodes all together are essential for the design of a Smart EM Environment where the electromagnetic field can be shaped through network control. More specifically, the fundamental question we pose is: ''Do we always need a full-fledged IAB node to relay signals, boost capacity, and enhance 5G mmWave coverage?''. For example, in a typical urban scenario, different parts of the city may require different types of Smart EM devices, depending on the performance objective of the target use case (see \S \ref{actors}). We refer to this concept as \textit{heterogeneous deployment}.    
%\begin{figure}[t]
%\centering
%\includegraphics[width=0.4\textwidth]{./pics/het_deployment}
%\caption{Heterogeneous deployment for the Smart EM environment.}
%\label{fig:het_dep}
%\end{figure}

\begin{figure}[t]
 \centering
 \subfigure[CDF of user received power]
   {\includegraphics[width=0.4\textwidth]{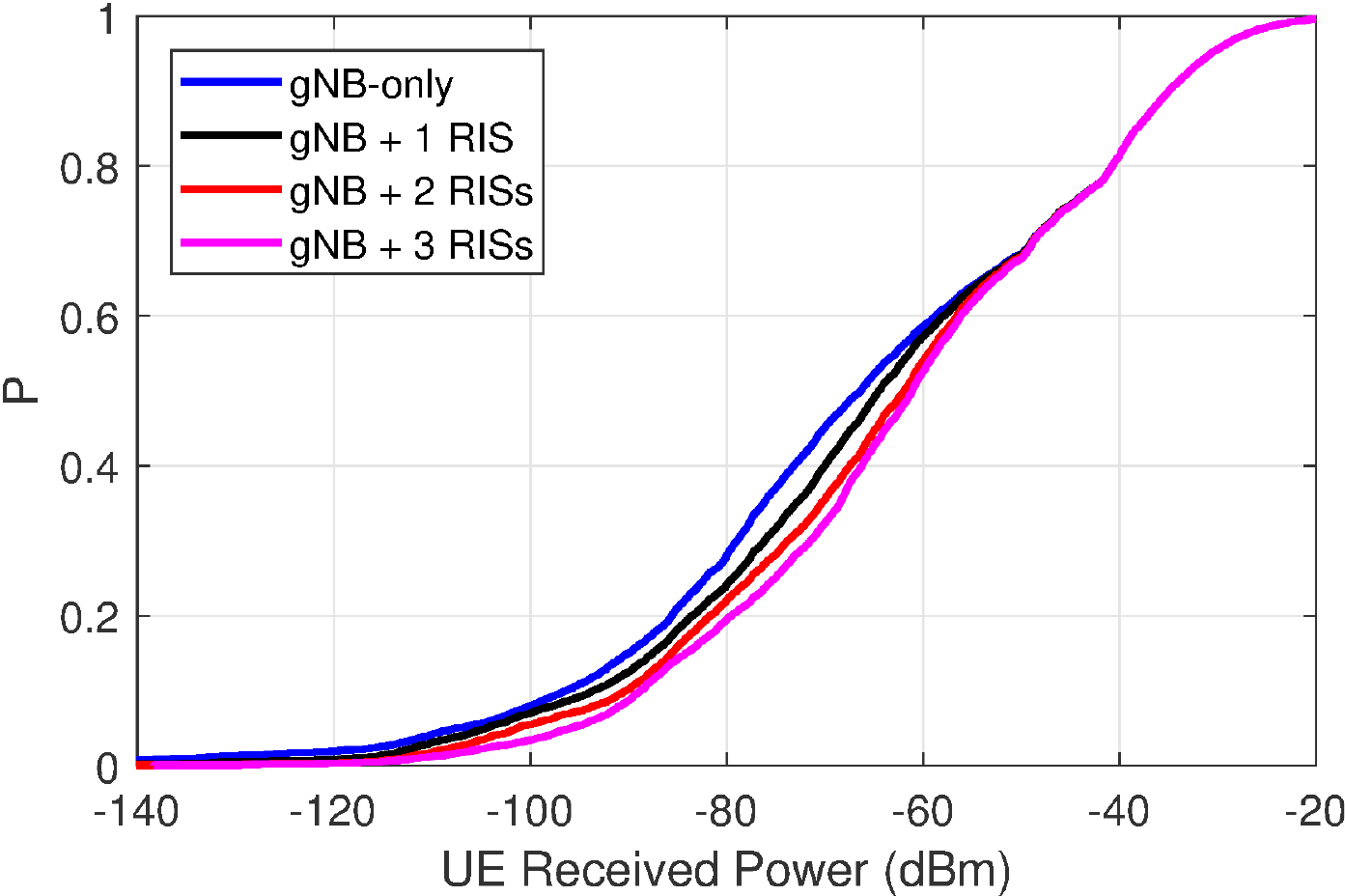} \label{fig:pwr}}
 \hspace{4mm}
 \subfigure[CDF of user capacity]
   {\includegraphics[width=0.4\textwidth]{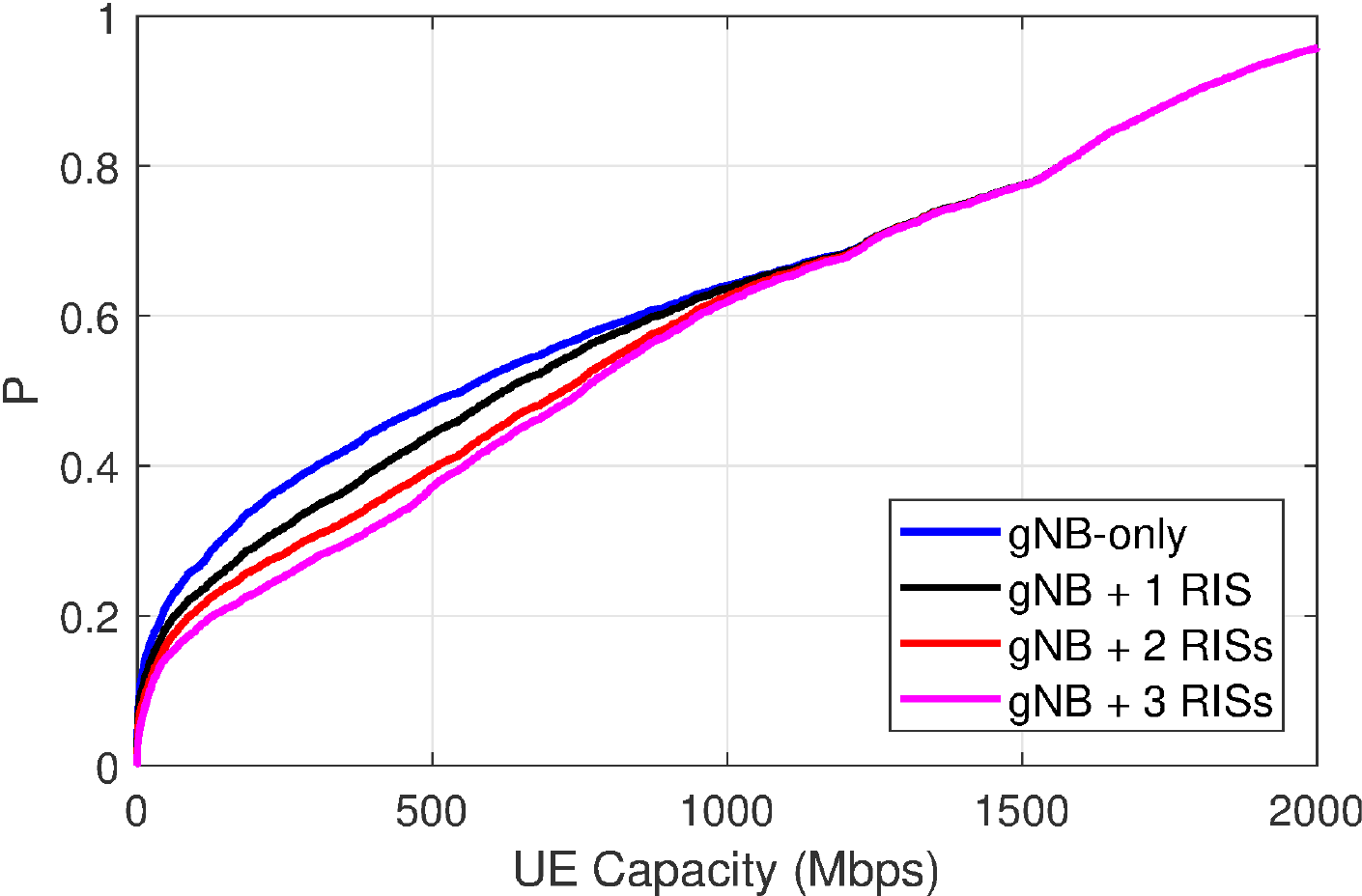} \label{fig:cap}}
 \caption{CDFs of user received power (a) and capacity (b) for the Hong Kong simulation scenario: comparison between gNB-only and gNB+RISs deployments.}
	\label{fig:HK_cdf}
 \end{figure}

The key idea behind the heterogeneous deployment is that while IAB nodes represent the most powerful tool to rapidly add 5G mmWave cells, they can be replaced in many cases by much cheaper Smart Repeaters, RISs, or Smart Skins without substantially affecting performance. To better investigate this aspect, we simulated via ray-tracing prediction a small area (roughly 100x100m$^2$) of Hong Kong city where we employed a 28GHz 5G gNB with 65dBm EIRP and 33dBi antenna gain installed at 10m height in a road intersection (see Fig.~\ref{fig:macro_only}). Several areas with very low received power can be observed, especially along the cross streets of the main road. Overall, the achieved 5G mmWave coverage is around 60\% assuming omnidirectional user equipment. IAB nodes in strategic positions could be installed to extend the coverage; however, one may wonder whether resorting to full-fledged nodes like IAB is a convenient choice. To answer this question, we supposed to install three 25x25cm$^2$ RISs at the minor road intersections, in line-of-sight with the gNB, as shown in Fig.~\ref{fig:MACRO_RIS}. A significant improvement in received power can be observed, which translates, approximately, into 20\% coverage increase. The impact of sequentially adding RIS 1, RIS 2, and RIS 3 as in Fig.~\ref{fig:MACRO_RIS} is reported in Fig.~\ref{fig:HK_cdf}, where the CDFs of UE received powers and achievable capacities is showed. The beneficial effect of deploying RISs is evident, especially for cell-edge users: received power is increased by more than 10dB while capacity achieves almost 2x improvement.  
     
%
%{\ed In order to analyze the impact of installing not only RISs, but also IAB nodes, Smart Repeaters and Smart Skins to form an heterogeneous deployment, we are currently carrying out statistical simulations using the 3GPP geometry-based stochastic channel model \cite{ref_3GPP_channel} in an urban scenario with hexagonal cell layout. Our objective is to show that, exploiting unconventional network planning optimization strategies, the heterogeneous deployment is able to provide substantial TCO reduction compared to an IAB-only deployment. More details about network planning of Smart EM environment are given in the next section.}

\section{System challenges} \label{system_challenges}

The heterogeneous deployment envisioned for the Smart EM Environment is able to deliver the expected benefits when proper network planning strategies are adopted and efficient coordination among the gNBs and the multiple IAB nodes, RISs and Smart Repeaters is achieved. This poses several challenges under the perspective of PHY layer and system design, as discussed in the following.

\subsection{Network planning}
In the heterogeneous deployment envisioned in \S \ref{het_deployment}, it is important to explore network planning strategies that take into account performance, cost, and energy-efficiency of each Smart EM node in order to improve the 5G mmWave coverage in realistic scenarios. In particular, robustness against obstacle blockages can be achieved by providing multiple connection alternatives to users, via relaying or reflections, in order to always make available an entry point for the mmWave access when random obstacles block some links. Smart EM nodes are capable of focusing an impinging radio wave towards arbitrary directions and this capability can be exploited at a network level to create a reflected radio path between a transmitter and a receiver. The tuple comprising a transmitter, a receiver and Smart EM device assisting the communications is known as Smart Radio Connection (SRC) \cite{pimrc_moro} and extends the well-known concept of Smart
Radio Environment \cite{Renzo2020}. SRCs can thus naturally provide dual-connectivity, namely connecting the UE through multiple radio links, which is
effective in reducing outages due to obstacle obstructions in mmWave access networks. However, a single, big nomadic obstacle could interrupt both the primary and reflected path if these present low angular separation, leading to an SRC outage and decreasing the solution's efficacy. A large angular separation between primary and reflected links can protect the SRC against obstructions and mitigate the well-known effect of self-blockage, caused by the human body of the device holder interrupting the line-of-sight. Link length represents another factor influencing the network resilience that can be controlled during the network planning phase. Indeed, longer radio connections naturally experience higher probabilities of being blocked by one or more obstacles. Independently on the link recovery procedure employed during network operation, a reliable mmWave RAN design coincides with a careful installation of gNB and Smart EM objects during the planning phase. 

\subsection{Impact of hardware impairments} 
An intensively studied approach for implementing RISs consists in resorting to the concept of metasurface, that is a digitally-tunable two-dimensional metamaterial composed of a wide number of reflecting elements, also referred to as meta-atoms, whose size is of the order of sub-wavelength of the operating frequency \cite{ref_11_jonathan}. Each meta-atom is equipped with reconfigurable circuits that can be tuned to modify the phase of the impinging electromagnetic wave. Therefore, the impact of the RIS onto the end-to-end baseband channel model can be modeled by a diagonal matrix where the non-zero entries represent the reflection coefficients. The phases of these coefficients constitute the tunable part, while the amplitudes are fixed and depend on the properties of the specific structure \cite{ref_1_jonathan}. 
On the other hand, Smart Repeaters can be seen as virtually omnidirectional amplify-and-forward nodes that are composed of a number of tunable antennas (e.g., phased arrays) able to synthesize transmit and receive beam patterns towards specific directions. Their channel model is typically captured by a full complex matrix where each entry depends on the choice of the transmit and receive beamforming vectors and the gain introduced in the amplification process. 

The possibility of tuning the phases (and when possible the amplitudes) over a continuous domain, while desirable under a performance point of view, would lead to a costly and not scalable manufacturing process. Practical implementations of RISs and Smart Repeaters privilege the use of a quantized set of phases and amplitudes to limit the number of controls for each radiating element. Typically adopted quantization levels are in the order of 1-4 bits, depending on the applications. The beamforming design should in principle explicitly account for this limitation, especially with implementations based on 1 or 2 bits, even if this policy may lead to very complex optimization routines. A more pragmatic approach consists in first deriving beamforming vectors with continuous phases and amplitudes, and then quantizing the obtained solutions to comply with the discrete sets.

\begin{figure}[t]
\centering
\includegraphics[width=0.44\textwidth]{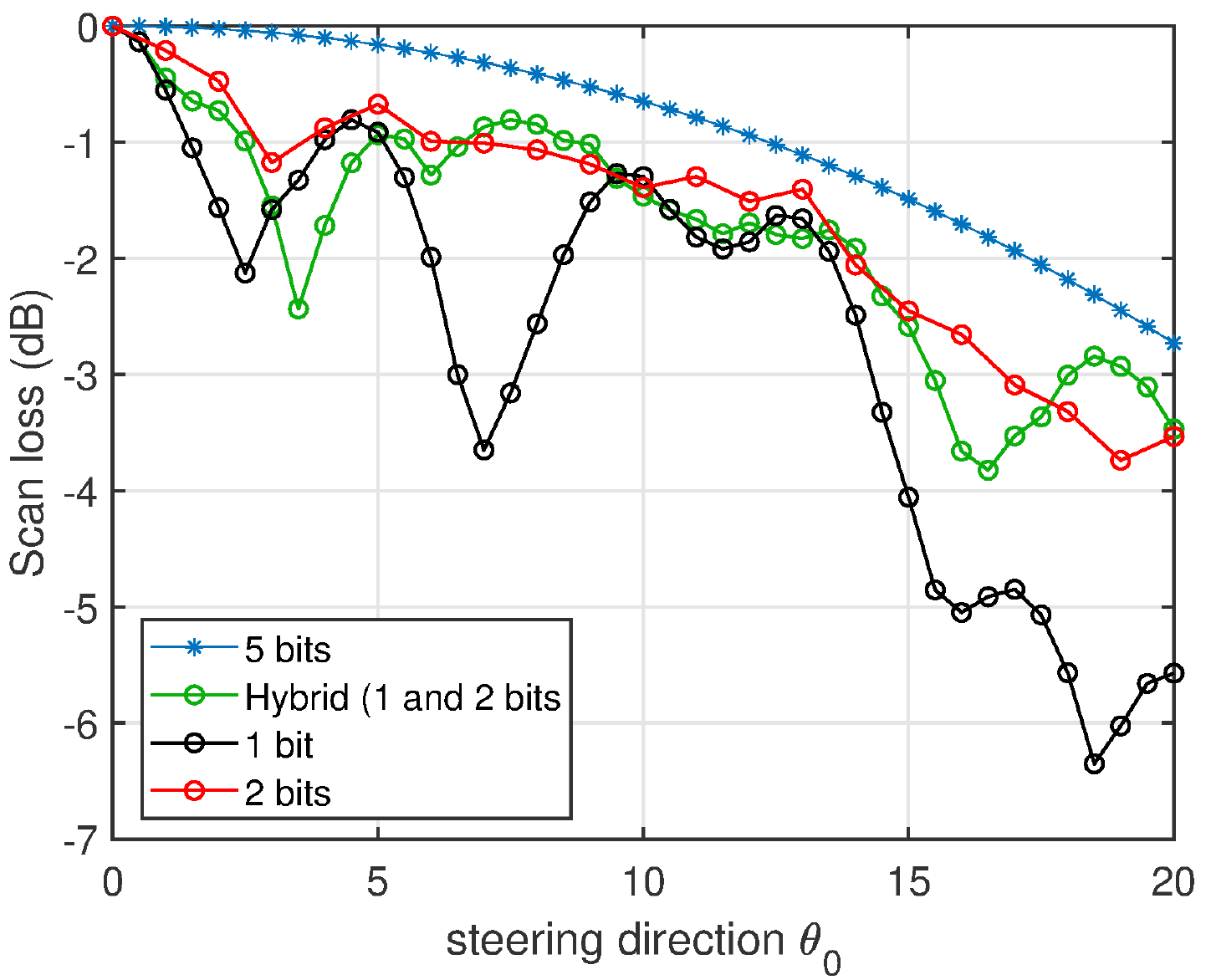}
\caption{Radiation pattern envelope for an antenna array where beam codewords have been optimized accounting for the available level of phase quantization (1 bits for all elements, 2 bits for all elements or hybrid 1 or 2 bits for the different elements).}
\label{fig:bit_quantization}
\end{figure}

The presence of phase quantization typically leads to the synthesis of patterns with lower directivity and higher side lobes with respect to the case with infinite resolution \cite{pattern_danilo}. To emphasize this concept, Fig.~\ref{fig:bit_quantization} shows the scan loss achieved when electronically pointing a phased array structure towards different elevation angles by employing beam codewords appropriately optimized with the respect to the admissible phases at each antenna element (the phased array is characterized by 8 directive elements along the vertical direction deployed with a mutual spacing of 1.5$\lambda$). Despite the optimized approach, the figure suggests a remarkable directivity loss when a 1-bit quantization is adopted.  

\subsection{CSI acquisition and beamforming design}
The transparent nature of the Smart EM Entities limits the possibility of estimating the single contributions of the gNB-to-UE relayed channels. To circumvent this issue, a practical solution relies on letting the RISs and Smart Repeaters operate with pre-designed codebooks of phase configurations, and consequently of beam patterns. During initial access, phase codewords should be able to synthesize beams wide enough to keep the users search time at an acceptable level. After the connection is established, convenient procedures based on channel state information (CSI) feedback form the UEs should be devised in order to increase the beam pattern directivity, combat poor path loss conditions, and maximize the achievable spectral efficiencies.

The codebook-based approach requires that the beam synthesis at the gNBs and the RISs or Smart Repeaters is tightly synchronized. Moreover, the gNB needs to be aware of the beam pattern that maximizes the energy transfer towards each Smart EM node deployed in the cell area. Since networks should be planned to guarantee LoS communications between each gNB and all the Smart EM entities, this issue can be handled by an initial network configuration phase. We remark that the codebook design for RIS nodes is site-specific, as it should depend jointly on both the source and the destination angles, and cannot be defined a priori. This implies that the codebooks to be adopted at the different RISs should be selected after the gNB-to-RIS relative azimuth and elevation angles have been accurately estimated. Finally, motivations related to cost and implementation issues will very likely lead to both RIS’s and Smart Relays with phase-only tunable elements. This opens room to the study and development of ad-hoc optimization routines for designing phase-only codebooks able to generate beam patterns with variable widths and shapes \cite{ref_20_jonathan, ref_21_jonathan, ref_22_jonathan}.

\subsection{Energy efficiency}\label{ef_section}
Beam switching of gNBs and Smart EM entities can in principle occur over different time scales, thus entailing diversified capacity versus energy efficiency trade-offs. In scenarios where high transmission gains are needed to optimize capacity, the Smart EM entities should tailor the reflection coefficients to maximize the directivity towards (or from) each intended user. This requires that the RISs and Smart Repeaters synthesize highly directive beams that need to be changed very frequently – typically in the order of one or multiple Transmission Time Intervals (TTIs), according to the scheduling needs. The vocation of low power consumption inherently characterizing the RISs (each element consumes typically less than 1 mW) can be pushed forward by letting them change their configuration at a much slower rate with respect to the TTI support -- e.g. in the order of one switch per units of second or even hour. In this case, the RIS should be configured to synthesize wider beam patterns to provide coverage to a target area that possibly encompasses multiple users. In this scenario, the capacity performance is traded with energy efficiency by setting the RIS circuits devoted to decode the control and synchronization signals in idle or even deep sleep modes for a great portion of the time. 

When the configuration switching time becomes large, the RIS appears more as an asynchronous network entity that can dynamically adjust the phase shifts of each atomic reflecting element based on the propagation environment learned through either appropriate control signals sent from the core network on a slow timescale basis, or even through periodic sensing carried out via the same passive array (when not reflecting) \cite{ref_5_jonathan}. In the latter case, the RIS should be equipped with very low-power receive RF chain(s) to enable the sensing capability for CSI acquisition, while the RIS controller should be able to select two different modes of operation: (\textit{i}) receiving mode for environment sensing; (\textit{ii}) reflecting mode for scattering the incident signals from the gNB.

\subsection{Users scheduling}
The scheduling problem in the advocated Smart EM Environment acquires an additional level of complexity with respect to traditional cellular systems, due to the need of efficiently managing the multiple passive or active relay nodes. After the beam management procedures have selected the candidate beam patterns to reach each user, both directly from the gNB or through relayed paths accounting for RISs and Smart Repeaters, the scheduler should assign the available Radio Resource Blocks by accounting for the hierarchical structure of the network architecture. Reports on the interference status by the UEs is expected to have a crucial role in this framework, to be managed in a centralized fashion. The resulting problem is non-convex in nature, and suitable heuristic methods should be devised that are able to scale well with the high problem dimensionality. The scheduler should be able to capitalize scenarios when the signal attenuation of the gNB-to-UE link is comparable to that of the RIS-relayed link by selecting the transmit beamforming of the gNB to strike a balance between the UEs and RISs angular directions. In this case, the phase shifts of the RIS need to be chosen in order to trigger a constructive sum at the receiver between the reflected signal and the signal coming directly from the gNB \cite{ref_25_jonathan}.

\section{Hardware challenges} \label{hardware_challenges}

All the actors envisioned for the Smart EM Environment that have been presented in \S \ref{actors} are requested to realize the promises of enhanced coverage and capacity while keeping the costs and the complexity of the network as low as possible. In order to successfully achieve this combination of high-performance, low-complexity and cost-effectiveness, the development of each Smart EM node faces several challenges in terms of both hardware design and technology, as discussed in the following.

\subsection{IAB node}
Being already standardized, the IAB node is probably the entity with the lowest degree of freedom for alternative implementations. There are currently two main research directions envisioned to improve the performance of the IAB node while reducing its impact on the network TCO: (\textit{i}) two-sector deployment and (\textit{ii}) Full-duplex (FD) operation. 

The first direction focuses on extending the Field of View (FoV) of the IAB antenna from a typical three-sectorial value of 120$^{\circ}$ to 180$^{\circ}$. In this way, the IAB node would be composed of only two antennas (instead of three) to provide omnidirectional coverage, hence reducing the overall hardware cost. The realization of a 180$^{\circ}$-scanning antenna system is challenged by system-level aspects, as the antenna has now to serve a larger FoV with same performance of the three-sector case, as well as hardware-level issues, since the development of an antenna with a 180$^{\circ}$ FoV requires innovative EM design techniques like conformal arrays \cite{ref_1_hw}, deflecting lenses\cite{ref_2_hw}, metasurfaces \cite{ref_3_hw}, reconfigurable pattern radiating elements \cite{ref_4_hw} and wide-pattern radiating elements suitable for phased arrays \cite{ref_5_hw}.

The second direction (FD operation) aims at improving drastically the node performance thanks to the fact that uplink and downlink transmissions can reuse the same radio resources \cite{ref_6_hw}. This allows to enhance capacity, reduce latency for both access and backhaul links and enable flexible and dynamic uplink/downlink resource adaptation according to the traffic pattern. Realizing efficient FD operation is challenging due to self-interference, clutter echo, and cross-link interference. Key issues for FD IAB nodes are the spatial isolation between antennas, the beam isolation between transmit and receive antenna patterns, the digital/analog self-interference mitigation techniques and suitable mechanisms to identify the optimal TX/RX beam pairing for user scheduling.

\begin{table*}[]
\centering 
\caption{Overview of main technologies for tunability and reconfigurability}
\begin{tabular}{m{2.8cm}|m{3cm}|m{2cm}|m{1.5cm}|m{2.5cm}|m{3cm}}
\hline \hline
Technology & \centering FoM$^*$/performance & \centering Cost & \centering Speed & \centering Max. pwr/IP3 & \centering Comment \tabularnewline \hline

PIN diode & \centering Up to 2THz FoM \cite{ref_1_table}, IL 0.6dB\@30GHz \cite{ref_2_table} & \centering $\sim$1\$ & \centering $\sim$10ns & \centering 1-10W P$_{max}$ &  \centering Mature technology, few suppliers at microwave and mmWave frequencies \tabularnewline \cline{1-6}

MEMS & \centering IL 0.5dB\@10GHz \cite{ref_3_table}  & \centering $\sim$10\$ & \centering $\sim$10$\mu$s & \centering 10W P$_{max}$ &  \centering Few suppliers in volume production \tabularnewline \cline{1-6}

SOI CMOS & \centering 2THz FoM \cite{ref_4_table}, 1dB\@10GHz \cite{ref_5_table}, 2dB\@30GHz \cite{ref_5_table}  & \centering $<$1\$ & \centering $\sim$50ns & \centering 50dBm IP3 &  \centering Suitable for RFIC integration, less as standalone device \tabularnewline \cline{1-6}

PHEMT (GaN) & \centering 0.5THz FoM \cite{ref_6_table} & \centering - & \centering $\sim$100ns & \centering $\sim$10W &  \centering Suitable for MMIC integration, less as standalone device \tabularnewline \cline{1-6}

Liquid crystals & \centering 120$^{\circ}$/dB \cite{ref_7_table} & \centering ''low'' & \centering $\sim$10ms & \centering $\sim$1W &  \centering Suitable for system integration, less as standalone device \tabularnewline \cline{1-6}

Phase change material & \centering 7THz FoM \cite{ref_8_table} & \centering ''low'' & \centering $\sim$1$\mu$s & \centering $\sim$1-10W &  \centering Data refer to GeTe. V02 still in early research phase \tabularnewline \cline{1-6}
 
Ferroelectric & \centering 100$^{\circ}$/dB \cite{ref_9_table} & \centering $\sim$1\$ & \centering $\sim$10ns & \centering $\sim$1-10W &  \centering Either standalone device or integrated into systems 	\tabularnewline \cline{1-6}

 \hline \hline
\multicolumn{6}{l}{\small *FoM=$1/(2\pi C_{off}R_{on})$} \\
\end{tabular}
\label{tab:technologies}
\end{table*}

\subsection{Smart Repeater}
Smart Repeaters are amplify-and-forward devices with a required end-to-end (E2E) gain of approximately 90dB for mmWave applications. This figure represents the actual bottleneck to Smart Repeater performance. In fact, in order to avoid detrimental oscillations, the isolation between the two antennas must be greater than the E2E gain. This has to be addressed providing adequate isolation margin, e.g., by achieving a proper tradeoff between antenna directivity, antenna spacing, and amplifier gain. However, being the EM environment where the Smart Repeater is deployed not fully known and time varying, an active stability check must be implemented, able to detect oscillation risks and take proper countermeasures (e.g., gain reduction, switch off, etc.). The use of echo/feedback cancelers can be envisioned here, but complexity and cost increase has to be taken into account.

\begin{figure}[t]
\centering
\includegraphics[width=0.46\textwidth]{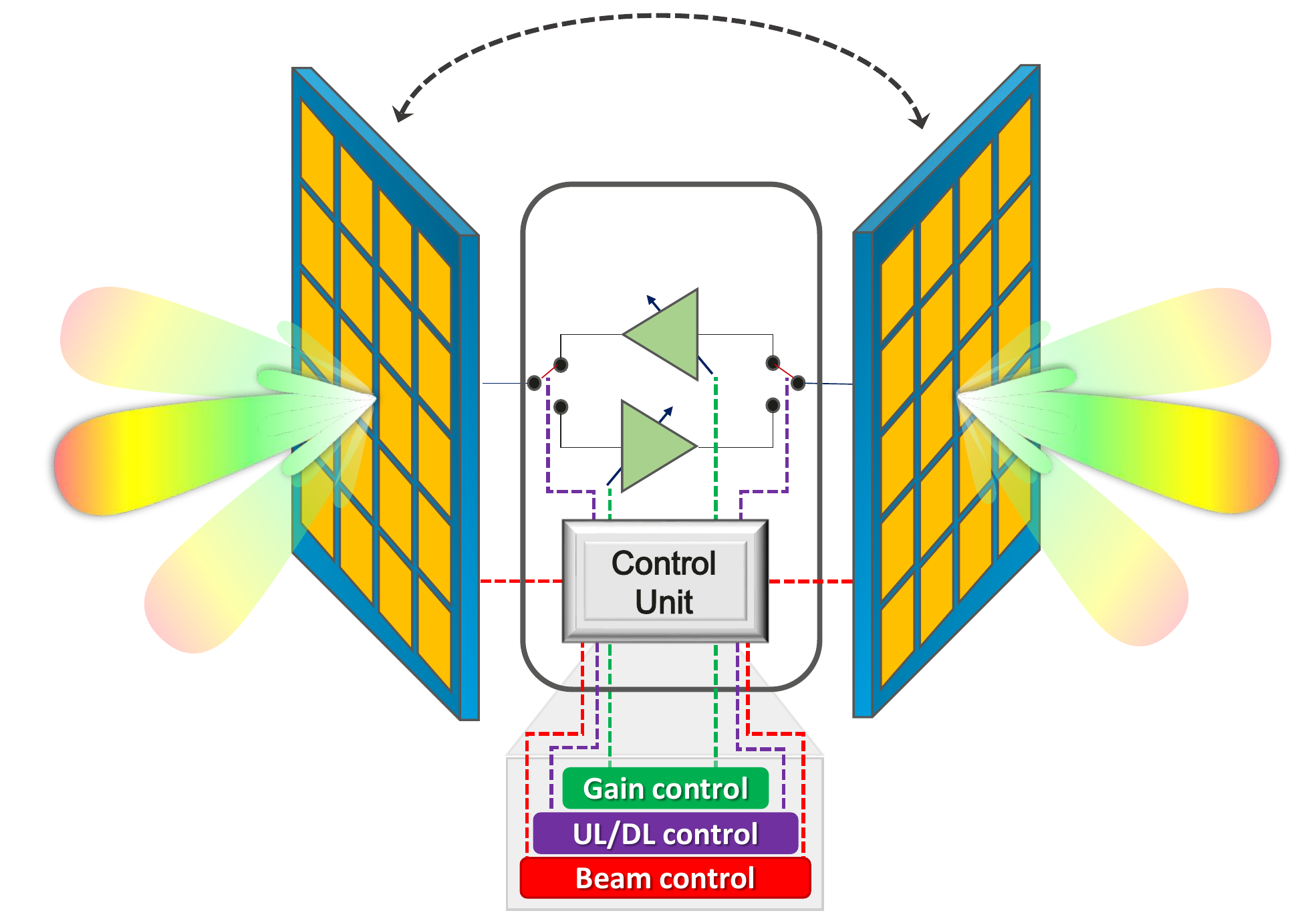}
\caption{Block diagram of the Smart Repeater comprising beamforming antennas, amplifiers, TDD switching and control unit.}
\label{fig:smart_repeater}
\end{figure}

From an implementation point of view, the Smart Repeater is a simple apparatus made of two beamforming antennas, one RF amplifier and a suitable control unit that, connected with the Donor gNB, manages beamsteering/beamforming algorithms, amplification control and uplink/downlink switching (see Fig.~\ref{fig:smart_repeater}). Considering a target EIRP of around 60dBm, the peak power required from the RF amplifier is in the order of a few Watts and can be readily obtained with current composite semiconductor technology. Uplink/downlink switches requires a high level of isolation that, with a smart circuit implementation, is still well within current technological capabilities. Beamforming antennas do not require the ultimate RF performance level of those used for Macro BS or IAB nodes, but rather have to be low cost and simple to produce. Conventional solutions rely on the use of a plurality of radiating elements that, alone or grouped into clusters (i.e., sub-arrays), are connected to a phase shifter and an amplification stage. Clustering in a phased array already goes in the direction of reducing the number of phase shifters in order to lower system complexity as well as to simplify heat dissipation. 

Within this context, a significant effort in our research activities is currently devoted to the development of low-cost and robust techniques to achieve phase-control capabilities and reliable yet continuous beam-steering with no need of conventional phase shifters. The first attempts to achieve antenna reconfiguration relying on PIN diodes instead of phase shifters date back to 1978, where preliminary studies for a PIN-based tunable dielectric waveguide were presented \cite{ref_A_hw}. In more recent studies, PIN diodes were used as switches to control the period of the grating, thus demonstrating beam steering in a number of discrete directions \cite{ref_C_hw}. At the moment, the interest of researchers is focused on the problem of reconfigurability applied to metasurfaces and obtained through the use of a number of different techniques, from PIN diodes to varactors, liquid crystals or other tunable materials \cite{ref_F_hw}. In a recent research activity in collaboration with the University of Siena and its spin off Wave-Up Srl, we have developed a solution based on a reconfigurable metasurface leaky-wave antenna, in which a network of properly biased PIN diodes is used to control the beam pointing angle in the elevation plane \cite{ref_G_hw,ref_H_hw}. In this design, the metasurface has been designed as composed of a plurality of sub-lambda radiating elements organized into different channels to create a leaky-wave effect. Each channel contributes to a beam whose pointing direction can be varied along the elevation plane; this possibility is ensured by the presence of separately-driven PIN diodes soldered across the radiating elements: by dynamically varying the modulation period through the voltage control of each diode, the antenna beam angle is steered accordingly, with a reconfiguration speed that falls in the order of tens or hundreds of nanoseconds (compatible with the latest communications technologies, such as 5G). One of the main advantages of such an approach is that the leaky-wave antenna can be realized in simple PCB technology, leading to a low-profile, cheap implementation suitable also for mass production. In addition to that, compared to traditional clustered phased-array topologies, this structure permits to extremely simplify the signal routing, as each channel has a unique feeding point; the considerable number of PIN diodes permits to achieve an arbitrary beam deflection (not constrained to a set of discrete angles), but, on the other hand, the use of a great number of active PIN diodes leads to relevant costs and some uncertainties in the actual yield of the fabrication process. It appears to be imperative to develop, possibly jointly with the manufacturers, a well-established and reliable circuital/electromagnetic model for the active devices (e.g., PIN diodes). In \cite{ref_I_hw}, an example of such an analysis for mm-Wave frequencies is presented. As a matter of fact, the performance of such components can be very variable according to the batch, the fabrication process and most importantly the kind of packaging, which often introduces parasitic capacitance and inductance that dominate the response. The unpredictability of the behavior of reconfigurable devices strongly affects the propagation constant of the surface wave along the metasurface. These considerations apply for RISs too (since are based on reconfigurable devices as well) and are of particular interest when dealing with mmWave systems. 

\subsection{RISs and Smart Skins}
As described in \S \ref{actors}, RISs and Smart Skins are surfaces able to reflect an incident beam to an angle which is different with respect to the incident one (non-specular reflection), or to provide additional focusing-defocusing and beam-shaping effects. The main challenge for such devices is to make them extremely inexpensive in order to create a real cost-effective solution that can be massively deployed. As an example, PCB technology, despite the extremely high optimization of the manufacturing processes achieved along decades of wide use, is still probably too expensive, considering the wide surfaces that are expected to be covered. The RIS poses additional challenges in finding a suitable reconfigurable technology able to cope with the performance and cost requirements. If we approach the design of a RIS as a collection of controllable unit cells whose size is $\lambda/2$ (or much less in case of a metasurface-based RIS), the cost of the controlling elements (PIN diodes, varactors, switches) must be in the order of a few cents of USD to keep overall price to an acceptable level. This is probably the most interesting research direction for the actual deployment of such solution: research on new, ultra-low-cost manufacturing technologies with the capability to integrate seamlessly the tuning element, or the development of a new concept of a virtually continuous (in space) tunable surface. On top of this, being reconfigurable, a RIS requires a control circuitry and a low capacity link with Donor gNB to synchronize its operation with the network; it would be highly desirable that RIS could be disconnected from the public power network and powered on its own, for example with batteries or by a suitable method of energy harvesting, being photovoltaic panels the most obvious one. This sets a limit of a few Watt for the RIS power consumption that is a severe additional constraint for the tunable technology used. As mentioned in \S \ref{ef_section}, switching speed, e.g., for semi-static and programmable surfaces, can make a significant difference in such an aspect.

\subsection{Technologies for tunability and recofigurability}
As discussed in the previous sections, Smart Repeaters and RISs need suitable reconfigurable/tunable technologies with a different price/performance ratio compared to the ones used in gNB or IAB apparatus. The remaining part of this section provides a high-level overview of the main technologies that are currently investigated for tunability and reconfigurability, alongside the more classical semiconductor-based ones. For the sake of completeness, some of them are listed in Table~\ref{tab:technologies}. Generally speaking, while semiconductor-based technologies (we consider MEMS along them, since their manufacturing process stems from semiconductor industry) are well developed and can offer a good level of performance, their price level is still not suitable for RIS targets. Material-based technologies (LCD, Ferroelectric, PCM, etc.) have a huge potential from this point of view, thanks to the simpler manufacturing process (often involving just a single deposition step) or the possibility to be seamlessly integrated into the RIS substrate, but they often lack from the performance point of view. A description of the main material-based technologies is provided below.

\subsubsection{Ferroelectric ceramic materials}
Ferroelectrics are special materials that show a non-null electric moment generating a spontaneous polarization which can be reversibly switched by an externally applied electric field. Ferroelectric materials (as Barium Strontium Titanate, BST) change their dielectric constant as a function of an external electric field and this makes them attractive for applications in electrically tunable microwave devices, such as tunable oscillators, phase shifters and varactors \cite{ref_1_laura,ref_2_laura,ref_3_laura}. One critical aspect for practical applications of ferroelectric materials is the reduction of the dielectric losses, which tend to be high when high performance is required in microwave devices. One strategy to reduce dielectric losses of BST ceramic thick or thin films could be to exploit the addition of oxides such as magnesium oxide (MgO) \cite{ref_1_laura}.  

\subsubsection{Phase change materials}
Phase Change Materials (PCMs) are receiving considerable attention as reliable candidates for next generation non-volatile memory devices, glass fibers, data storage media, high density optical recording applications, microchips, and reprogrammable photonic circuits \cite{ref_4_laura,ref_5_laura,ref_6_laura}.  These materials are characterized by the capability to change from amorphous to crystalline phase, in order to display different electrical features. The transformation between states could be triggered by a thermal excitation generated by an optical or electrical pulse. Germanium-Antimony-Tellurium (GST) alloys are the most widespread PCMs due to the fast reversibility of the amorphous-to-crystalline state transition and the resulting pronounced alternation of different optical and electrical properties \cite{ref_4_laura,ref_5_laura}. GST-PCMs are fascinating materials to be used in electrical and optical sectors, but some issues could arise in real applications, such as power consumption, amorphous state instability after repeated cycling, excessive resistivity contrast and materials degradation. Among PCMs materials, it is worth mentioning also Vanadium Oxides, which show semiconductor to metal/metal to insulator transitions \cite{ref_7_laura}. Vanadium trioxide (V$_2$O$_3$), in particular, shows a remarkable change in resistivity, but its applicability is limited by the low transition temperature. For this reason, vanadium dioxide (VO$_2$), that has a transition temperature above room temperature, is more frequently exploited. 

\subsubsection{Nematic Liquid Crystals}
Ordered fluid mesophases, commonly called Liquid Crystals, simultaneously possess liquid-like (fluidity) and solid-like (molecular order) properties whose unique coexistence gives rise to many interesting properties. Nematic Liquid Crystals (NLCs) are elongated molecules composed of a number of aromatic rings and two end groups, at least one of which  is polar \cite{ref_7_laura} and enables their easy reorientation by using a DC or low frequency electric field. Low losses at high frequencies and low bias voltage have recently made the use of NLCs interesting in the devices at microwave frequency range.

\section{Conclusion} \label{conclusion}
A Heterogeneous Smart Electromagnetic Environment for mmWave communications has been presented, describing in detail the new network nodes that are part of it and the related progresses in the activities of both the standardization bodies and the industrial players. Within this framework, a system-level study has been carried out showing some preliminary results of the heterogeneous deployment in terms of both coverage and capacity and highlighting the major system challenges that need to be faced. Lastly, some of the most recent technological directions both for the hardware development and for the research on tunable devices have been provided, giving useful insights from the industrial point of view.

%\vspace{-2mm}
\bibliographystyle{IEEEtran}
\bibliography{IEEEfull,mybib}

\vfill

% That's all folks!
\end{document}